\documentclass[aps,preprint,tightenlines,floatfix,superscriptaddress,nofootinbib,showpacs,]{revtex4}
\usepackage{graphicx}
\usepackage{multirow}
\usepackage{amsmath}
\usepackage{hyperref}
\usepackage{color}
\usepackage{appendix}
\def\beq{\begin{equation}}
\def\eeq{\end{equation}}
\def\bea{\begin{eqnarray}}
\def\eea{\end{eqnarray}}
\def\nn{\nonumber}

\def\roughly#1{\mathrel{\raise.3ex\hbox
{$#1$\kern-.75em\lower1ex\hbox{$\sim$}}}}

\def\tbar{{\overline{t}}}
\def\bbar{{\overline{b}}}
\def\sbar{{\overline{s}}}
\def\cbar{{\overline{c}}}
\def\ellbar{{\overline{\ell}}}
\def\nubar{{\overline{\nu}}}
\def\tbbc{t \to b \bbar c}
\def\ggprocess{gg\to t\tbar\to\left(b\bbar c\right)
    \left(\bbar\ell\nubar\right)}

\def\ggantiprocess{gg\to t\tbar\to\left(b \bar \ell \nu \right)
    \left(b \bbar \bar c \right)}
\def\qqbarprocess{q\overline{q}\to t\tbar\to\left(b\bbar c\right)
    \left(\bbar\ell\nubar\right)}   
\def\tm#1{\texttt{TM-{#1}}}
\def\tme#1{\texttt{TME-{#1}}}
\def\ex#1{\texttt{EX-{#1}}}

\def\madg{\textsc{MadGraph5}}
\def\feynr{\textsc{FeynRules}}
\def\ACP{A_{\rm TP}}

\begin{document}
\bibliographystyle{apsrev}

\preprint{\vbox {\hbox{UdeM-GPP-TH-15-243}}}

\vspace*{2cm}

\title{\boldmath Measuring CP-Violating Observables \\ in Rare Top Decays at the LHC}

\def\umontreal{\affiliation{\it Physique des Particules, Universit\'e
    de Montr\'eal, \\ C.P. 6128, succ.\ centre-ville, Montr\'eal, QC,
    Canada H3C 3J7}}
\def\tayloru{\affiliation{\it Physics and Engineering Department,
    Taylor University, \\ 236 West Reade Ave., Upland, IN 46989, USA}}
\def\laplata{\affiliation{\it IFLP, CONICET -- Dpto. de F\'{\i}sica,
    Universidad Nacional de La Plata, C.C. 67, 1900 La Plata,
    Argentina}}
\def\hri{\affiliation{\it Harish-Chandra Research Institute, Chhatnag Road, Jhunsi, 
    Allahabad - 211019, India}}

\umontreal
\tayloru
\laplata
\hri

\author{Pratishruti Saha}
\email{pratishrutisaha@hri.res.in}
\umontreal
\hri

\author{Ken Kiers}
\email{knkiers@taylor.edu}
\tayloru

\author{Bhubanjyoti Bhattacharya}
\email{bhujyo@lps.umontreal.ca}
\umontreal

\author{David London}
\email{london@lps.umontreal.ca}
\umontreal

\author{Alejandro Szynkman}
\email{szynkman@fisica.unlp.edu.ar}
\laplata

\author{Jordan Melendez}
\email{melendez.27@osu.edu}
\altaffiliation{Current address:
  Department of Physics, The Ohio State University,
  191 West Woodruff Ave.,
  Columbus, OH 43210, USA.}
\tayloru

\date{\today}

\begin{abstract}
In this paper we consider CP-violating new-physics contributions to the decay
$t \to b \bar b c$. We examine the prospects for detecting such new physics at 
the LHC, which requires studying the process 
$gg \to t (\to b \bar b c) \bar t (\to \bar b \ell \bar \nu)$. We find two 
observables that can be used to reveal the presence of CP-violating new 
physics in $t \to b \bar b c$. They are (i) the partial-rate asymmetry and 
(ii) the triple-product correlations involving the momenta of various 
particles associated with the interaction.  A Monte Carlo analysis is 
performed to determine how well these observables can be used to detect the 
presence of new physics, and to measure its parameters. We find that there 
is little difficulty in extracting the value of the relevant new-physics 
parameter from the partial-rate asymmetry. For the triple-product 
correlations, we test multiple strategies that can be used for the 
extraction of the corresponding combination of new-physics parameters.
\end{abstract}


\maketitle

\section{Introduction}
\label{sec:intro}

It is widely believed that physics beyond the Standard Model (SM) must
exist.  However, to date, no evidence of this new physics (NP) has
been found.  It appears that the energy scale of the NP is larger than
was hoped for, or that its manifestation is subtler than envisioned.
Over the years many models of NP have been proposed, and a number of
these feature the top quark in a central role~\cite{topNPmodels}.  
Being particularly heavy, with a mass near the electroweak scale, 
the top quark may well be sensitive to NP interactions that do not much 
affect other SM particles. On the other hand, top observables such as 
total cross-section~\cite{topcsec}, decay width~\cite{topwidth}, 
differential cross-sections~\cite{topdifferential}, etc. appear to be in good
agreement with the corresponding SM predictions. Significant NP contributions 
may therefore exist only in processes that are suppressed in the SM. 
One such process is the decay $\tbbc$. The SM rate for this process is very 
small as it involves the Cabibbo-Kobayashi-Maskawa (CKM) element 
$V_{cb}$ ($\sim$ 0.04).

NP contributions to $\tbbc$ were studied in Ref.~\cite{kklrsw}, and
several observables that can reveal the presence of NP were found.
This decay can be studied at the LHC, which is essentially a top-quark
factory.  However, single-top production is rather suppressed at the
LHC~\cite{singletprod}, so that it is difficult to isolate $\tbbc$
experimentally and analyze it on its own. Instead, one considers 
$t \tbar$ pairs that are produced predominantly through gluon fusion: 
$gg \to t \tbar$. 
The $t$ and $\tbar$ then decay into a pair of $b$-jets along with other
hadronic and/or leptonic final states. In order to study $\tbbc$, it
is useful to consider the semi-leptonic channel $gg \to t (\to b \bbar
c) \tbar (\to \bbar\ell\nubar)$ where the charge of the lepton may be
used to ascertain that it is the $t$ that is undergoing the rare
decay.

In Refs.~\cite{tNPLHC1,tNPLHC2} a detailed numerical simulation of $gg
\to t (\to b \bbar c) \tbar (\to \bbar\ell\nubar)$ was performed to
examine how well NP parameters can be determined at the LHC when it
operates at 14 TeV. This analysis focused on CP-conserving NP. In
the present paper, we examine the possibilities for detecting
CP-violating NP and measuring its parameters. In Ref.~\cite{kklrsw} it
was shown that there are two observables that are sensitive to
CP violation in $\tbbc$ -- the partial-rate asymmetry and the triple
product. In the full process, $gg \to t (\to b \bbar c) \tbar (\to
\bbar\ell\nubar)$, one has these same two observables.  We examine
each of these observables separately.  For the partial-rate asymmetry,
the analysis is straightforward. However, as we will see, for the
triple product it is more involved.

We begin in Sec.~\ref{sec:NPintbbc} by describing the effective
Lagrangian describing NP contributions to $\tbbc$ and outlining the
calculation of the differential cross section for $\ggprocess$.  In
Sec.~\ref{sec:observables} we define CP-violating observables in
$\ggprocess$. Included here are the analytic expressions for the
partial-rate asymmetry and the triple product in this process. In
Sec.~\ref{sec:numerical} we detail the numerical simulations performed
to determine how well the CP-odd NP parameter combinations can be
extracted from measurements of the partial-rate asymmetry, the triple
product and related observables. We discuss the feasibility of
measuring $\ggprocess$ in Sec.~\ref{sec:feasibility}.  We conclude in
Sec.~\ref{sec:conclusions}.


\section{\texorpdfstring{\boldmath New Physics Contributions to $t$ decay}{New Physics Contributions to t decay}}
\label{sec:NPintbbc}

\subsection{\texorpdfstring{\boldmath $\tbbc$: effective Lagrangian}{Effective Lagrangian}}
\label{sec:eff_lag}

The decay $t\to b \overline{b}c$ can have contributions coming from
the SM ($t\to bW^+\to b\overline{b}c$) and from various NP sources.
We parameterize the NP contributions via an effective Lagrangian, as
was done in Refs.~\cite{kklrsw,tNPLHC1,tNPLHC2}: we set
${\cal L}_{\mbox{\scriptsize eff}} = {\cal L}_{\mbox{\scriptsize eff}}^V +
{\cal L}_{\mbox{\scriptsize eff}}^S + {\cal L}_{\mbox{\scriptsize eff}}^T$, 
with
\begin{eqnarray}
  {\cal L}_{\mbox{\scriptsize eff}}^V & = & 4\sqrt{2}G_F V_{cb}V_{tb}
       \left\{
    X_{LL}^V\,\bbar\gamma_\mu P_L t \,
       \overline{c}\gamma^\mu P_L b
   + X_{LR}^V\,\bbar\gamma_\mu P_L t \,
       \overline{c}\gamma^\mu P_R b
\right.\nonumber\\
& & \hskip2.2truecm \left.
   +~X_{RL}^V\,\bbar\gamma_\mu P_R t \,
       \overline{c}\gamma^\mu P_L b
   + X_{RR}^V\,\bbar\gamma_\mu P_R t \,
       \overline{c}\gamma^\mu P_R b
\right\}+ \mbox{h.c.}, 
\label{eq:eff1}\\
&&\nonumber\\
  {\cal L}_{\mbox{\scriptsize eff}}^S & = & 4\sqrt{2}G_F V_{cb}V_{tb}
\left\{
     X_{LL}^S\,\bbar P_L t \,\overline{c} P_L b
   + X_{LR}^S\,\bbar P_L t \,\overline{c} P_R b
\right. \nonumber\\
& & \hskip2.2truecm \left.
   +~X_{RL}^S\,\bbar P_R t \,\overline{c} P_L b
   + X_{RR}^S\,\bbar P_R t \,\overline{c} P_R b
\right\}+\mbox{h.c.,} 
\label{eq:eff2}\\
&&\nonumber\\
  {\cal L}_{\mbox{\scriptsize eff}}^T & = & 4\sqrt{2}G_F V_{cb}V_{tb}
\left\{
     X^T_{LL} \overline{b}\sigma^{\mu\nu}P_L t \,
     \overline{c}\sigma_{\mu\nu}P_L b 
\right. \nonumber\\
& & \hskip2.2truecm \left.     +~X^{T}_{RR}
\bbar\sigma^{\mu\nu}P_R t \,
     \overline{c}\sigma_{\mu\nu} P_R b
\right\}+\mbox{h.c.}
\label{eq:eff3}
\end{eqnarray}
The colour indices in the above expressions are assumed to contract
in the same manner as those in the SM; Ref.~\cite{kklrsw} contains
an analysis of the case in which the indices contract differently
than in the SM.

The dimensionless NP parameters $X^I_{AB}$ in
Eqs.~(\ref{eq:eff1})-(\ref{eq:eff3}) may be assumed to be $O(1)$.
Under this assumption, the NP contributions to $t\to b\overline{b}c$
can be of the same order as that coming from the SM.  For this reason,
when analyzing possible NP effects it is important to consider not
just the SM-NP interference terms, but also the NP-NP pieces.  In this
paper we focus specifically on CP-violating effects, which can arise
when the $X^I_{AB}$ contain weak phases.  Throughout this work we
ignore strong phases related to NP contributions, since these are
negligible~\cite{DatLon}.  There is a strong phase related
to the $W$ resonance in the SM contribution to the decay; this phase
plays an important role in the partial-rate asymmetry (see
Sec.~\ref{sec:PRA}).


\subsection{\texorpdfstring{\boldmath Differential cross section for $\ggprocess$}{Differential cross section}}
\label{sec:diffcsec}

The differential cross section for $\ggprocess$ was worked out in
Ref.~\cite{tNPLHC1}. In this section we summarize the procedure; the
results can be found in Appendix~\ref{sec:app1}. The full details are 
given in Ref.~\cite{tNPLHC1}.

\begin{figure}[!htbp]
\begin{center}
\resizebox{4in}{!}{\includegraphics*{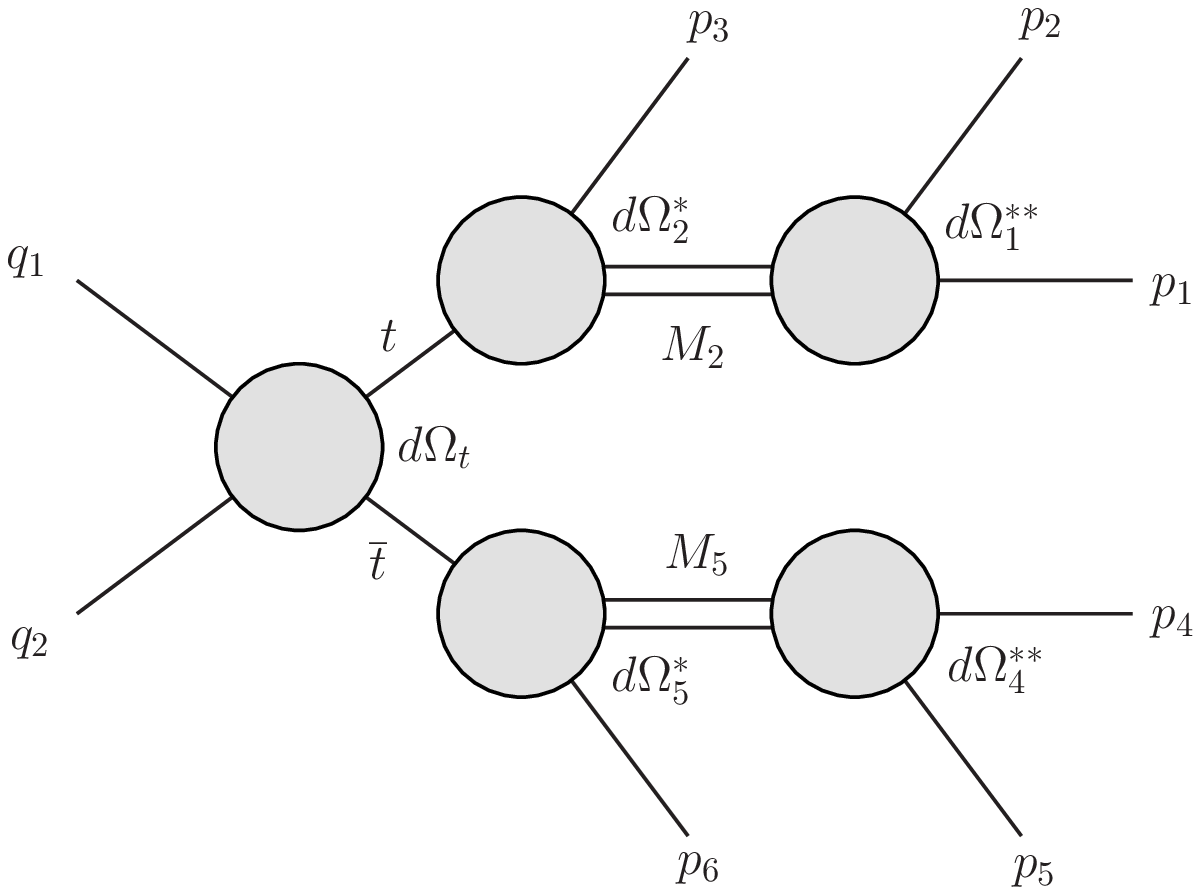}}
\caption{Kinematics for 
  $\ggprocess$~\cite{ggprocesskinematics}. The definitions
  of the various angles and invariant masses are identical
  to those given in Ref.~\cite{tNPLHC1}; these definitions are
  reproduced here for convenience.  $\Omega_1^{**}$ denotes
  the direction of $\vec{p}_1^{~**}$ in the rest frame of $M_2$,
  relative to the direction of $\vec{p}_1^{~*}+\vec{p}_2^{~*}$, where
  $M_2^2 = (p_1+p_2)^2$.  Similarly, $\Omega_2^*$ denotes the
  direction of $\left(\vec{p}_1^{~*}+\vec{p}_2^{~*}\right)$ in the $t$
  rest frame, relative to the direction of $\vec{p}_t$ in the $t\tbar$
  rest frame.  $\Omega_t$ denotes the direction of $\vec{p}_t$
  relative to $\vec{q}_1$, also in the $t\tbar$ rest frame.  The solid
  angles $\Omega_4^{**}$ and $\Omega_5^{*}$ are defined analogously to
  $\Omega_1^{**}$ and $\Omega_2^{*}$, respectively, and $M_5^2 =
  (p_4+p_5)^2$.  In this work we take $p_1=p_c$, $p_2=p_b$,
  $p_3=p_{\overline{b}_1}$, $p_4=p_{\overline{\nu}}$,
  $p_5=p_{\overline{b}_2}$ and $p_6=p_\ell$.}
\label{fig:kinematics}
\end{center}
\end{figure}

The kinematics of the process $\ggprocess$ is
represented in Fig.~\ref{fig:kinematics}.
As described in Ref.~\cite{tNPLHC1},
the six-body phase space may be decomposed into five solid angles and 
four invariant masses.  
  Note that Fig.~\ref{fig:kinematics} represents only the 
  kinematics of $\ggprocess$ -- it is not a Feynman diagram. 
  Thus, $M_5^2$ does not necessarily correspond to the $W^-$ resonance in 
  the $\tbar$ decay, and $M_2^2$ does not necessarily correspond to the 
  $W^+$ resonance in the SM part of the $t$ decay. 
  Rather, $p_1$, $p_2$ and $p_3$ are the
  momenta of the $b$, $\bbar$ and $c$ quarks in $\tbbc$, with all
  permutations being allowed.  Assuming that the
$t$ and $\tbar$ quarks are on-shell before decaying, two of the
invariant-mass degrees of freedom can be eliminated. The solid angles
$d\Omega_1^{**}$, $d\Omega_2^{*}$, $d\Omega_4^{**}$, $d\Omega_5^{*}$
and $d\Omega_t$ in Fig.~\ref{fig:kinematics} are defined in five different rest frames,
with the $*$ and $**$ superscripts
indicating that these angles are defined in reference frames that are,
respectively, one and two boosts away from the $t\tbar$ rest
frame. The invariant masses $M_2$ and $M_5$ are defined via
$M_2^2 = \left(p_1+p_2\right)^2$ and $M_5^2 =
\left(p_4+p_5\right)^2$.  The differential cross section is a
complicated function of the various momenta
  [see Eqs.~(\ref{eq:dsigma2})-(\ref{eq:dlambda})]; these momenta
  may in turn be related back to the solid angles and invariant masses
  via boosts and rotations.

The approximate analytical expression for the differential cross
section for $gg \to t \tbar \to (b \bbar c)(\bbar\ell\nubar)$ was
derived making several simplifying assumptions:
\begin{enumerate}

\item We considered only the $gg$ initial state, ignoring
  $q\overline{q}$ initial states.
\item We ignored the parton distribution functions (PDFs) for the
  initial gluons and worked in the rest frame of the initial $gg$
  pair.  In the actual experiment, the initial gluons have a wide
  range of momenta, and the lab frame is generally different than the
  $gg$ rest frame for a given event.
\item We considered the final state $\overline{b}$'s to be
  ``distinguishable,'' when in fact they are identical particles.

\end{enumerate}
In addition to the above simplifications, we also set the masses
of the light quarks and the charged lepton to zero.
The analytical expressions for the differential cross section and
integrated cross section for $\ggprocess$ are given in Appendix~\ref{sec:app1}. 
At first glance, it might appear that the above assumptions would have
rendered these expressions almost completely useless.  On the contrary,
however, we have found that these expressions provide crucial insights
into the actual physical process and serve as a useful starting point
for a more robust numerical treatment of the problem.

In Refs.~\cite{tNPLHC1,tNPLHC2} we focused primarily on CP-even
observables. In the present work we turn our attention to CP-odd
observables. We proceed in the same manner as we did in
Refs.~\cite{tNPLHC1,tNPLHC2}, working first from theoretical
expressions derived under various simplifying assumptions, and turning
later to a more robust numerical treatment.



\section{\texorpdfstring{\boldmath CP-violating Observables in $\ggprocess$}{CP-violating observables}}
\label{sec:observables}

A perusal of the general expressions for the differential and
integrated cross sections for $gg \to t (\to b \bbar c) \tbar (\to
\bbar\ell\nubar)$ shows that that there are two CP-odd
combinations\footnote{This statement is true in the limit that the
  light quarks are taken to be massless.  There are other CP-odd
  combinations of NP parameters that show up in $t\to b \overline{b}c$
  if we relax this
  assumption, but they are suppressed by $\sim O(m_b/m_t)$.} of NP
parameters that can be probed in this process, namely Im($X^{V*}_{LL}$)
[Eq.~(\ref{eq:sigggttbar_int})] and
Im($X^T_{LL}X^{S*}_{LL}+X^T_{RR}X^{S*}_{RR}$) [Eq.~(\ref{eq:BTP2})].
These same two parameter combinations were analyzed in
Ref.~\cite{kklrsw}, although the notation in that paper was somewhat
different. In addition, for
Im($X^T_{LL}X^{S*}_{LL}+X^T_{RR}X^{S*}_{RR}$) it was assumed there
that the spin of the top quark could be measured (obviously a
simplifying assumption).  In the present context, the correlations
between the pair-produced $t$ and $\tbar$ effectively allow us to gain
access to the spin of the top.

The CP-violating observables that will allow experimentalists to
measure the above CP-odd NP parameter combinations are as follows.
Im($X^{V*}_{LL}$) is probed using the partial-rate asymmetry, while
Im($X^T_{LL}X^{S*}_{LL}+X^T_{RR}X^{S*}_{RR}$) appears in triple
products and can be probed in several ways. In the following subsections 
we describe each of these observables in turn. Note that the analytic 
expressions, wherever quoted, have been derived with the simplifying 
assumptions discussed above.

\subsection{Partial rate asymmetry}
\label{sec:PRA}

The simplest CP-odd observable may be obtained by comparing the cross
section for the process $\ggprocess$ to that for the conjugate process 
$\ggantiprocess$. Now,
CP-violating effects can only arise as a result of the interference of
two amplitudes. Furthermore, all signals of direct CP violation, such
as the partial-rate asymmetry (PRA), are proportional to the CP-odd
quantity $\sin\phi \sin\delta$, where $\phi$ and $\delta$ are
respectively the weak-phase and strong-phase differences between the
two amplitudes. As noted in Sec.~\ref{sec:eff_lag}, the NP strong
phases are negligible, so $\delta$ is due entirely to the SM
$W$-mediated amplitude. Furthermore, the weak phase 
must arise entirely from NP since the SM weak phase is $\simeq 0$. 
Therefore, the PRA is due to SM-NP interference. The only NP contribution 
that interferes with the SM is the $(V-A)\times(V-A)$ term in the effective 
Lagrangian. As a result, the PRA is proportional to the width of the $W$ and 
to Im($X^{V*}_{LL}$).

Normalizing to the sum of the cross sections, the PRA can be written
\begin{eqnarray}
{\cal A} = \frac{\sigma-\overline{\sigma}}{\sigma+\overline{\sigma}}
\simeq \frac{1}{\cal R}
\frac{4\Gamma_W}{m_W}\mbox{Im}\left(X^{V*}_{LL}\right),
\label{eq:PRA}
\end{eqnarray}
where 
\begin{eqnarray}
{\cal R} =
\frac{\sigma+\overline{\sigma}}{2\sigma_{\mbox{\scriptsize
    SM}}} = 
    1+ \frac{3 G_F m_t^2}
     {4\sqrt{2}\pi^2 \left(1-\zeta_W^2\right)^2
      \left(1+2\zeta_W^2\right)}
    \sum_{i,\sigma} \hat{A}_i^\sigma
\label{eq:R}
\end{eqnarray}
with $\zeta_W \equiv m_W/m_t$ and $\hat A_i^{\sigma}$ being combinations 
of various $X^I_{AB}$, as defined in Eq.~(\ref{eq:Ahatdefs}).

While the presence of the ratio $\Gamma_W/m_W$ in Eq.~(\ref{eq:PRA})
leads to a suppression of the PRA, it is still possible to obtain an
asymmetry whose magnitude is in excess of 10\% \cite{kklrsw}. And,
despite this suppression, the PRA still offers several advantages. The
foremost among these is that it is relatively straightforward to
measure, since it does not require a detailed kinematical analysis or
the determination of angles in various rest frames.  One simply counts
the number of events for the $t$ decay in this channel and compares that 
to the number of events for the $\tbar$ decay in the analogous channel.  
In fact, since the PRA does not
require the presence of correlations between the pair-produced $t$ and
$\tbar$, we needn't be so restrictive regarding the decay mode of the
``other'' particle.  That is, we could just as well compare the width
for $gg \to t (\to b \bbar c) \tbar (\to \mbox{everything})$ to that
for $gg \to \tbar (\to \bbar b \cbar) t (\to \mbox{everything})$ in
order to increase statistics (assuming, of course, that the process and 
conjugate process could still be distinguished without tagging on the 
charge of the lepton). We present the numerical results for benchmark NP 
scenarios in Sec.~\ref{sec:PRA_sim}.


\subsection{Triple product}
\label{sec:TP}

In decay processes with two contributing amplitudes $A$ and $B$, the
square of the total amplitude may contain interference terms of the
form Im($AB^*$)[$\vec v_1 \cdot (\vec v_2 \times \vec v_3)$], where each
$v_i$ is a spin or a momentum. These triple products (TPs) are odd under
time reversal (T) and hence, by the CPT theorem, also constitute
potential signals of CP violation. Now,
\beq
{\rm Im}(AB^*) 
= 
|A||B| ( \sin\phi \cos\delta + \cos\phi \sin\delta ) ~,
\label{ImAB*}
\eeq
where $\phi$ and $\delta$ are respectively the weak-phase and
strong-phase differences between $A$ and $B$. The first term is
CP-odd, while the second is CP-even, so that the TP is not by itself a
signal of CP violation (this is due to the fact that T is an
anti-unitary operator). On the other hand, the TP in the CP-conjugate
process is proportional to
\beq
{\rm Im}(AB^*)_{CP-conj} 
= 
|A||B| ( - \sin\phi \cos\delta + \cos\phi \sin\delta ) ~.
\label{ImAB*conj}\eeq
Combining ${\rm Im}(AB^*)$ and ${\rm Im}(AB^*)_{CP-conj}$ allows one
to isolate the CP-odd piece proportional to $\sin\phi \cos\delta$. That is, 
as with direct CP violation (the PRA), in order to obtain a CP-violating 
signal, one must compare the TP in the process with that in the CP-conjugate
process. However, in contrast to direct CP violation, no strong-phase
difference between the interfering amplitudes is required in order to
obtain a non-vanishing CP-violating signal 
(i.e. $\delta$ can be 0). It is interesting to note that, if the strong 
phase difference is indeed negligible ($\delta \approx 0$), then the 
CP-even term (proportional to $\cos\phi \sin\delta$) is approximately zero, 
which then makes the TP a signal of CP-violation by itself.

In Ref.~\cite{kklrsw}, it was shown that, in the presence of NP, a TP
of the form 
$\vec{s}_t\cdot\left(\vec{p}_{\overline{b}}\times \vec{p}_c\right)$ 
can be generated in the decay $\tbbc$. Here $\vec{s}_t$ denotes the spin of 
the $t$, and $\vec{p}_i$ is the momentum of the particle $i$ coming from the 
decay of the $t$. Since the top decays, one might try to gain access to the 
top's spin via correlations with the momenta of its decay products. Such an 
approach cannot give access to a quantity such as 
$\vec{s}_t\cdot\left(\vec{p}_{\overline{b}}\times \vec{p}_c\right)$, 
however, since the three momenta $\vec{p}_i$ ($i=b,{\bar b},c$) are not 
independent. The problem can be circumvented by using the fact that, in 
$t \tbar$ production, the spins of the $t$ and the $\tbar$ are statistically 
correlated~\cite{spincorrexpt}. As $\vec{s}_\tbar$ is related to the momenta 
of the decay products of the $\tbar$, the TP in $\tbbc$ can be rewritten as a
TP involving three final-state momenta of the full process, 
$gg \to t (\to b \bbar c) \tbar (\to \bbar\ell\nubar)$, and this does not
vanish. In practice, this is implemented by introducing the $t\tbar$ 
spin-correlation coefficient:
\begin{equation}
\label{eq:spincorrcoeff}
\kappa_{t \tbar} 
\quad = \quad 
\dfrac{   \sigma_{\uparrow\uparrow} + \sigma_{\downarrow\downarrow}
        - \sigma_{\uparrow\downarrow} - \sigma_{\downarrow\uparrow}
      }
      {   \sigma_{\uparrow\uparrow} + \sigma_{\downarrow\downarrow}
        + \sigma_{\uparrow\downarrow} + \sigma_{\downarrow\uparrow}
      } ~.
\end{equation}
Here, $\uparrow$ and $\downarrow$ denote the alignment of the spins of
the top and antitop with respect to the chosen spin-quantization
axis. As noted above, $\vec{s}_t$ is related to the momenta, or
angular distribution, of the $t$ decay products, and similarly for
$\vec{s}_\tbar$. The TP in $gg \to t (\to b \bbar c) \tbar (\to
\bbar\ell\nubar)$ then involves the angular correlation between the
decay products of the two particles.  As is evident in Eq.~(\ref{eq:BTP2}),
there are also triple-product terms relating the initial-state gluons
and the decay products of the top.

As was noted above, the CP-odd combination of NP parameters that shows up in 
the triple-product terms is Im($X^T_{LL}X^{S*}_{LL}+X^T_{RR}X^{S*}_{RR}$) 
(see Appendix~\ref{sec:csec_diff}). That is, the TP is due to 
NP-NP interference. Furthermore, since the NP strong phases are negligible, 
$\delta = 0$ in Eqs.~(\ref{ImAB*}) and (\ref{ImAB*conj}). 
Hence, following the discussion below Eq.~(\ref{ImAB*conj}), the TP by itself 
is a signal of CP-violation in $\ggprocess$. In the sub-sections that follow 
we identify observables that can be used to isolate the TP and quantify the 
resulting CP-violation.


\subsubsection{Angular Distributions}
\label{sec:angdist}

The first observable is the double differential distribution relative to the 
angles $\theta_5^*$ and $\phi_1^{**}$. Of these, $\theta_5^*$ is related to 
the lepton polar angle in the $\overline{t}$ rest frame, while $\phi_1^{**}$ 
is an azimuthal angle in the $b$-$c$ rest frame.\footnote{$\theta_5^*$ is 
 defined in the $\overline{t}$ rest frame.  In this frame, we define the $z$ 
 axis to be the direction of the boost from the $t\overline{t}$ rest frame to 
 the $\overline{t}$ rest frame.  $\theta_5^*$ is the angle between the $z$ 
 axis and the $\overline{b}_2 \nu$ center of mass direction in this frame.  
 $\phi_1^{**}$ is defined in the $bc$ center of mass frame.  We define the 
 $z$ axis in that frame to be the direction of the boost from the $t$ rest 
 frame to the $bc$ rest frame. The $\overline{t}$ momentum in this frame 
 is taken to be in the $x-z$ plane, with its $x$-component being non-negative.
 This completely defines the coordinate system in which $\phi_1^{**}$ is then
 calculated as the usual azimuthal angle of the $c$ quark's momentum.}
Integrating the differential cross section 
over all phase-space variables except for these two angles yields
\begin{eqnarray}
  \frac{d\sigma}{d\!\cos\theta_5^* \,d\phi_1^{**}}
  &=& \frac{\sigma_{\mbox{\scriptsize SM}}}{4\pi} \Bigg\{ 1 +
    \frac{4\Gamma_W}{m_W}\mbox{Im}\left(X^{V*}_{LL}\right) 
    \; + \; \frac{3 G_F m_t^2}{4\sqrt{2}\pi^2 \left(1-\zeta_W^2\right)^2
      \left(1+2\zeta_W^2\right)} \Bigg(\sum_{i,\sigma}\hat{A}{_i^\sigma} \nonumber \\
    && \mspace{60mu} +\; \frac{2\pi^2\kappa(r)}{35}\Big[\cos\theta_5^*\cos\phi_1^{**}
         \left(\hat{A}_b^--\hat{A}_b^+-\hat{A}_c^-+\hat{A}_c^+\right)\nonumber\\
 && \mspace{115mu} + \; 16 \cos\theta_5^*\sin\phi_1^{**} \mbox{Im}\!\left[X^T_{LL}X^{S*}_{LL}+X^T_{RR}X^{S*}_{RR}\right]
         \Big]\Bigg)\Bigg\},
\label{eq:TP1}
\end{eqnarray}
where $\sigma_{\rm SM}$ is given by Eq.~(\ref{eq:sigmaSM}), $\zeta_W = m_W/m_t$ and
\begin{eqnarray}
  \kappa(r) &&= \;\; \frac{\left(-31 r^4 + 37 r^2 -66\right)r 
    -2\left(r^6-17r^4 + 33 r^2-33\right)\tanh^{-1}\left(r\right)}
        {r^2\left[\left(31 r^2 -59\right)r 
          +2\left(r^4-18r^2 + 33\right)\tanh^{-1}\left(r\right)\right]}
~,
\end{eqnarray}
with $r=\sqrt{1-4m_t^2/Q^2}$ and $Q\equiv p_t+p_{\overline{t}}$.
Note that $\kappa(r)$, as defined above,
differs from $\kappa_{t\overline{t}}$ in Eq.~(\ref{eq:spincorrcoeff}) by an overall
sign; also, $\kappa_{t\overline{t}}$ is averaged over energies.

Equation~(\ref{eq:TP1}) contains both CP-even and CP-odd terms.  The
part of the expression that is proportional to $\kappa(r)$ arises from
$t \tbar$ spin correlations~\cite{Soni}.  These terms disappear upon integration
over the angles $\theta_5^*$ and $\phi_1^{**}$, as one might expect.  

The term proportional to $\cos\theta_5^*\cos\phi_1^{**}$ in Eq.~(\ref{eq:TP1})
is sensitive to the CP-even combination of NP parameters
$(\hat{A}_b^--\hat{A}_b^+-\hat{A}_c^-+\hat{A}_c^+)$.  This combination
is distinct from the NP parameter combinations that arise in the
observables described in Refs.~\cite{tNPLHC1} and \cite{tNPLHC2}.  
Thus, although
our emphasis in the present work is on CP-odd observables, we note
that Eq.~(\ref{eq:TP1}) leads to a complementary approach to measuring
CP-even combinations of NP parameters.  The term that is of primary
interest to us in this work is the one proportional to
$\cos\theta_5^*\sin\phi_1^{**}$.  This term arises from the
triple-product terms in $\tbbc$ and contains the CP-odd NP parameter
combination Im($X^T_{LL}X^{S*}_{LL}+X^T_{RR}X^{S*}_{RR}$).
The value of Im($X^T_{LL}X^{S*}_{LL}+X^T_{RR}X^{S*}_{RR}$)
can be extracted directly by fitting the angular distribution
in Eq.~(\ref{eq:TP1}) using the template method developed in Ref.~\cite{tNPLHC2}.
We perform such a fit here for a few benchmark NP scenarios.
The details of the fitting procedure, our choice of templates, as well 
as the results are presented in 
Sec.\ref{sec:angdist_sim}.


\subsubsection{\texorpdfstring{$\langle \cos\theta_5^* \sin\phi_1^{**}\rangle$}{\textless ~cos(theta\_5)~sin(phi\_1)~ \textgreater}}
\label{sec:expecval}

Equation~(\ref{eq:TP1}) is also suggestive of a second observable that can be 
used to extract the value of Im($X^T_{LL}X^{S*}_{LL}+X^T_{RR}X^{S*}_{RR}$).
This is the expectation value of $\cos\theta_5^* \sin\phi_1^{**}$.
Taking into account the overall normalization, we find
\begin{eqnarray}
  \langle \cos\theta_5^* \sin\phi_1^{**}\rangle \; = \; 
  \frac{\sigma_{\mbox{\scriptsize
      SM}}}{\sigma} \left(\frac{2\sqrt{2}G_F m_t^2 \kappa(r)}
          {35 \left(1-\zeta_W^2\right)^2
      \left(1+2\zeta_W^2\right)}\right)
        \mbox{Im}\!\left[X^T_{LL}X^{S*}_{LL}+X^T_{RR}X^{S*}_{RR}\right] \mspace{20mu}
\label{eq:expec_val}
\end{eqnarray}
for a fixed value of the gluon energy.
For $pp$ collisions, one convolutes over parton distribution functions. 
This can be incorporated in an approximate way by making the replacement 
$\kappa(r)\to\langle \kappa(r)\rangle$, with $\langle \kappa(r) \rangle$ 
being measured over the events included in the analysis.  
From Eq.~(\ref{eq:expec_val}) we see that 
$\langle \cos\theta_5^* \sin\phi_1^{**}\rangle
\left(\sigma/\sigma_{\mbox{\scriptsize SM}}\right)$ 
as a function of 
$\mbox{Im}\!\left[X^T_{LL}X^{S*}_{LL}+X^T_{RR}X^{S*}_{RR}\right]$
is expected to be a straight line passing through the origin.
However, as mentioned earlier, this expression has been derived under 
the simplifying assumptions discussed in Sec.\ref{sec:diffcsec}.
To see how well this relation holds up in a more realistic scenario, 
we perform a Monte Carlo simulation where we generate data sets with 
different choices for 
$\mbox{Im}\!\left[X^T_{LL}X^{S*}_{LL}+X^T_{RR}X^{S*}_{RR}\right]$.
The results are detailed in Sec.\ref{sec:expecval_sim}.


\subsubsection{\texorpdfstring{$\ACP$}{A\_TP}}
\label{sec:ACP}

The third observable that can be used to capture the effect of the TP
is the quantity $\ACP$, which we define as 
\begin{equation}
\ACP 
\quad = \quad 
\dfrac{N[\epsilon(p_b,p_{\bar b},p_c,p_{\ell}) > 0] \;\; - \;\; 
       N[\epsilon(p_b,p_{\bar b},p_c,p_{\ell}) < 0]}
      {N[\epsilon(p_b,p_{\bar b},p_c,p_{\ell}) > 0] \;\; + \;\; 
       N[\epsilon(p_b,p_{\bar b},p_c,p_{\ell}) < 0]} \,,
       \label{eq:ATP}
\end{equation}

\noindent
where $\epsilon(p_b,p_{\bar b},p_c,p_{\ell}) 
= 
\epsilon_{\mu\nu\rho\lambda} \,
p_b^{\mu} \, p_{\bar b}^{\nu} \, p_c^{\rho} \, p_{\ell}^{\lambda}$ with 
$\epsilon^{0123} = +1$. 

Equation (\ref{eq:BTP2}) contains several terms of the type
$\epsilon(q_i,q_j,q_k,q_l)$, where the $q_i$ are momenta or combinations 
of momenta of the initial and/or final state particles. 
Of these, 
one expects that $\epsilon(p_b,p_{\bar b},p_c,p_{\ell})$ 
would be quite amenable to experimental measurement, as it only involves 
the measurement of the 4-momenta of the final state $b$, $\bar b$, $c$ and 
lepton. Moreover, it does not require the reconstruction of any special 
frames of reference and can be measured in the lab frame itself. 
Note that the measurement of $\ACP$ would not lead to the measurement of 
Im($X^T_{LL}X^{S*}_{LL}+X^T_{RR}X^{S*}_{RR}$) as such.
Nevertheless, a non-zero value of $\ACP$ would be a smoking gun signal of the
presence of CP-violating NP.  Furthermore, upon measurement of
a non-zero signal, it is expected that detailed numerical simulations could be used to
constrain the value of Im($X^T_{LL}X^{S*}_{LL}+X^T_{RR}X^{S*}_{RR}$).

Once again, we perform a Monte Carlo simulation for certain benchmark NP 
scenarios, the results of which are presented in Sec.~\ref{sec:ACP_sim}.




\section{Numerical Results}
\label{sec:numerical}

In this section we present the results of the numerical simulations to which 
we have alluded earlier. 
All the analytic expressions presented hitherto were obtained under the 
simplifying assumptions discussed in Sec.\ref{sec:diffcsec}.
However, for our numerical analysis we return to a more realistic
treatment. To be specific :
\begin{enumerate}
\item We include the contribution from $q\overline{q}$ initial states.
      This can be calculated in a manner similar to that 
      used for obtaining the $gg$ contribution (see Appendix~\ref{sec:app2}). 
      At the LHC, it gives only a sub-dominant contribution 
      ($\sim$ 10\%-15\%). Nevertheless, it is interesting to note that 
      the structure of distributions such as the one in Eq.~(\ref{eq:TP1})
      remains the same.
      In fact, the only change appears in the expressions for 
      $\sigma_{\rm SM}$
      and $\kappa(r)$. This, of course, is expected because in 
      Eq.~(\ref{eq:TP1}), these are the only two pieces that depend on the 
      $t\bar t$ production mechanism. The rest relates exclusively to the 
      dynamics of the decay.
\item We incorporate PDFs appropriately for the initial state partons.
\item We implement a procedure to distinguish between the identical 
      $\bar b$'s in the final state and identify ``correctly'' the $\bar b$
      coming from the $t$ decay. To do this, we construct the quantities 
      $m_1^2 = (p_b + p_c + p_{\bar b_1})^2$ and 
      $m_2^2 = (p_b + p_c + p_{\bar b_2})^2$. 
      If both $m_1$ and $m_2$ lie within $m_t \pm 15\Gamma_t$, the event is
      discarded. Otherwise, the $\bar b_i$ that leads to a smaller value of
      $|m_i - m_t|$ is assumed to come from the $t$ decay. For the
      conjugate process 
      ($p p \to t (\to b\bar \ell\nu) \tbar(\to \bbar b \cbar) )$, 
      a similar criterion is applied to the $b$'s.  The result is a loss 
      of $\sim$ 20\% of the events for both process and conjugate process.
\end{enumerate}
In addition, in generating the simulation data, we allow
  the light quarks and the charged lepton to have non-zero masses.
The event samples have been generated using \madg~\cite{MG5} in conjuction 
with \feynr~\cite{FR}. We consider a few benchmark NP scenarios to test 
the efficacy of the observables discussed above. $\sqrt{s}$ is taken to be 
14 TeV and CTEQ6L~\cite{CTEQ} parton distribution functions are used 
with both factorization and renormalization scales set to $m_t$ = 172 GeV. 
The integrated luminosity corresponds to $10^5$ SM events of the type 
$pp \to t \tbar \to \left(b\bbar c\right) \left(\bbar\ell\nubar\right)$,
which is expected to be achieved by the year 2030~\cite{SteveMyers}.

\subsection{Partial rate asymmetry - Results}
\label{sec:PRA_sim}

In this section we consider two benchmark NP scenarios\footnote{See Table~\ref{tab:NPcases} in
Appendix~\ref{sec:app3} for details of the choices made for the $X^I_{AB}$.}, which we label
\ex{A} and \ex{B}.
It is clear from Table~\ref{tab:PRA} that even when 
Im($X^{V*}_{LL}$) $\sim {\cal O}$(1), ${\cal A}$ can be fairly large ($\sim$ 10\%).
We also use Eqs.~(\ref{eq:PRA}) and (\ref{eq:R}) to extract the value of 
Im($X^{V*}_{LL}$) from the ``data''. 
Note that in real life we would 
have no a priori knowledge of the $\hat A^{\sigma}_i$'s. 
Therefore ${\cal R}$ would have to be calculated in terms of the observed
$\sigma$ and $\bar \sigma$ and the expected $\sigma_{\rm SM}$. 
As can be seen from the last column of Table~\ref{tab:PRA}, the values of 
Im($X^{V*}_{LL}$) are recovered quite accurately.
As expected, the PRA provides a simple and effective way to capture the 
effect of CP-violation in $\tbbc$.

\begin{center}
\begin{table}[ht]
\renewcommand{\arraystretch}{1.5}
\renewcommand{\tabcolsep}{0.2cm}
\begin{tabular}{|c|c|c|c|}
\hline 
\multirow{2}{*}{Model} & 
\multirow{2}{*}{Input Im($X^{V*}_{LL}$)} & 
\multirow{2}{*}{${\cal A} = \dfrac{N - \overline N}{N + \overline N}$} & 
\multirow{2}{*}{Extracted value of Im($X^{V*}_{LL}$)} \\
 & 
 & 
 &
 \\
\hline \hline
\texttt{SM} & 
0.0 & 
$-$0.002 $\pm$ 0.002 & 
$-$0.01 $\pm$ 0.02 \\
\hline
\ex{A} & 
$-$3.0 & 
$-$0.117 $\pm$ 0.001 &
$-$2.97 $\pm$ 0.03 \\
\hline
\ex{B} & 
$-$2.0 &  
$-$0.060 $\pm$ 0.001 &
$-$1.97 $\pm$ 0.04 \\
\hline
\end{tabular}
\caption{Partial rate asymmetries and recovered NP parameter values for the 
SM and two NP models.}
\label{tab:PRA}
\end{table}
\end{center}


\subsection{Triple product - Results}
\label{sec:TP_results}

\subsubsection{Angular Distributions}
\label{sec:angdist_sim}

In an experiment, a distribution of the type of Eq.~(\ref{eq:TP1})
would be measured as a 2-D histogram, {\tt D}. 
We see that the RHS of Eq.~(\ref{eq:TP1}) can be expressed as the sum of 
five terms -- one term independent of NP parameters and four terms dependent 
on Im($X^{V*}_{LL}$), $\sum \hat A_i^{\sigma}$, 
$(\hat{A}_b^--\hat{A}_b^+-\hat{A}_c^-+\hat{A}_c^+)$ and 
Im($X^T_{LL}X^{S*}_{LL}+X^T_{RR}X^{S*}_{RR}$), respectively.
Using {\madg}, and with appropriate choices for the $X^I_{AB}$, one can generate this 
angular distribution for a case where only one of the NP 
parameter combinations is non-zero and all others are zero. 
This can be done in turn for each of the four combinations.
In addition, there would be the case corresponding to the SM, where all the 
NP parameter-combinations are zero. These histograms form the templates 
that we label \tm0, \tm1, \tm2, \tm3, \tm4. 
Now the measured histogram {\tt D}, in which the
NP parameters take arbitrary, unknown values, can be expressed as a linear 
combination of the templates \tm{i} with appropriate weights; i.e.,
\begin{equation}
\mathtt{
D 
\quad = \quad 
w_0\tm0 \;+\; w_1\tm1 \;+\; w_2\tm2 \;+\; w_3\tm3 \;+\; w_4\tm4 \,.
}
\label{eq:D}
\end{equation}
The weights $\mathtt{w_i}$ can be determined through a simple fitting
procedure such as $\chi^2$ minimization and be used to extract the values 
of Im($X^{V*}_{LL}$), $\sum \hat A_i^{\sigma}$, 
$(\hat{A}_b^--\hat{A}_b^+-\hat{A}_c^-+\hat{A}_c^+)$ and 
Im($X^T_{LL}X^{S*}_{LL}+X^T_{RR}X^{S*}_{RR}$) encoded in the data histogram 
{\tt D}.

While the idea is simple, there are a few subtleties that must be 
taken care of during its implementation:
\begin{itemize}
 \item[-] First, the parameter inputs are provided in terms of $X^I_{AB}$. 
          By doing so it is possible to ensure that only one
          $\hat A^{\sigma}_i$ is non-zero at a time. However, it can still
          lead to overlapping contributions in the templates that we are 
          interested in. For example, a non-zero input for $\hat A^{+}_c$ 
          makes $\sum \hat A_i^{\sigma}$ as well as 
          $(\hat{A}_b^--\hat{A}_b^+-\hat{A}_c^-+\hat{A}_c^+)$ non-zero 
          simultaneously. These kinds of overlaps need to be removed.
          How we do this can be seen in Table~\ref{tab:templates}.
       
 \item[-] Second, a $\chi^2$ fit, by construction, can only distinguish 
          between terms with different angular structure. Equation~(\ref{eq:TP1}) 
          contains three terms with no angular dependence: the SM term, 
          the term proportional to Im($X^{V*}_{LL}$) and the term 
          proportional to $\sum \hat A_i^{\sigma}$.
          The fit is not capable of identifying the contributions coming from 
          these three pieces separately. To circumvent this problem, we
          fix the SM contribution to 1.0 and assume that Im($X^{V*}_{LL}$)
          could be fixed to the value 
          obtained by measuring the PRA. Thereafter we extract the values of  
          $\sum \hat A_i^{\sigma}$, 
          $(\hat{A}_b^--\hat{A}_b^+-\hat{A}_c^-+\hat{A}_c^+)$ and 
          Im($X^T_{LL}X^{S*}_{LL}+X^T_{RR}X^{S*}_{RR}$).         
          
 \item[-] Third, the template distributions must be subjected to the same 
          selection criteria, cuts, etc. as the data. 
\end{itemize}

We implement this fitting algorithm for two NP 
scenarios\footnote{See Table~\ref{tab:NPcases} in Appendix~\ref{sec:app3}
for details of the choices made for the $X^I_{AB}$.}. Once again, we generate 
``pseudo-data'' using {\madg}. 
Here we have specifically chosen NP scenarios where Im($X^{V*}_{LL}$) = 0,
to demonstrate the efficacy of the procedure in the ``best-case'' scenario 
when Im($X^{V*}_{LL}$) is actually 0. In the case of non-zero 
Im($X^{V*}_{LL}$), the value of Im($X^{V*}_{LL}$) estimated from the PRA is 
an input to the fit. This is also true of the observables discussed 
in Refs.~\cite{tNPLHC1} and \cite{tNPLHC2} where we focused primarily on
CP-even observables and set Im($X^{V*}_{LL}$) to zero in much 
of the analysis. Hence, while attempting to extract NP parameters in $\tbbc$, 
the first task would be to measure the PRA and the value of Im($X^{V*}_{LL}$).

The results of the fit are presented in Table~\ref{tab:fit}. It can be seen 
that, despite all the complications, the extracted values lie relatively 
close to the input values, although some of the fit values are several
standard deviations away from the corresponding inputs.  More importantly, 
the presence of NP, of both CP-conserving and CP-violating varieties, 
is firmly established. 

\begin{center}
\begin{table}[ht]
\renewcommand{\arraystretch}{1.5}
\renewcommand{\tabcolsep}{0.2cm}
\begin{tabular}{|c|c|c|c|}
\hline 
Template & 
$X^I_{AB}$ & 
$\hat A_i^{\sigma}$ & 
Surviving Contribution \\
\hline \hline
\tm{0} & 
All $X^I_{AB}$ = 0 & 
All $\hat A_i^{\sigma}$ = 0 & 
SM \\
\hline
\tm{1} & 
\multicolumn{2}{c|}{\tme5 $-$ \tme1} & 
Im($X^{V*}_{LL}$) \\
\hline
\tm{2} & 
\multicolumn{2}{c|}{\tme2 $+$ \tme3} & 
$\sum \hat A^{\sigma}_i$ \\
\hline
\tm{3} & 
\multicolumn{2}{c|}{\tme2 $-$ \tme3} & 
$(\hat{A}_b^--\hat{A}_b^+-\hat{A}_c^-+\hat{A}_c^+)$ \\
\hline
\tm{4} & 
\multicolumn{2}{c|}{\tme{4} $-$ \tme1 $-$ \tme2} & 
Im($X^T_{LL}X^{S*}_{LL}+X^T_{RR}X^{S*}_{RR}$) \\
\hline \hline
\tme{1} & 
Re($X^V_{LL}$) $\neq$ 0& 
$\hat A_{\bbar}^{+}$ $\neq$ 0; all other $\hat A_i^{\sigma}$ = 0 &
$\hat A_{\bbar}^{+}$ \\
\hline
\tme{2} & 
$X^V_{LR}$ $\neq$ 0 &  
$\hat A_{c}^{+}$ $\neq$ 0; all other $\hat A_i^{\sigma}$ = 0 &
$\hat A_{c}^{+}$ \\
\hline
\tme{3} & 
$X^V_{RL}$ $\neq$ 0 & 
$\hat A_{c}^{-}$
$\neq$ 0; all other $\hat A_i^{\sigma}$ = 0 & 
$\hat A_{c}^{-}$ \\
\hline
\multirow{2}{*}{\tme{4}} & 
Re($X^S_{LL}$) $\neq$ 0 ; & 
$\hat A_{\bbar}^{+}$, $\hat A_{c}^{+}$ $\neq$ 0 ; &
$\hat A_{\bbar}^{+}$, $\hat A_{c}^{+}$, \\
 & 
Im($X^T_{LL}$) $\neq$ 0 \hspace{3pt} & 
all other $\hat A_i^{\sigma}$ = 0 &
Im($X^T_{LL}X^{S*}_{LL}+X^T_{RR}X^{S*}_{RR}$) \\
\hline
\tme{5} & 
Im($X^V_{LL}$) $\neq$ 0 & 
$\hat A_{\bbar}^{+}$ $\neq$ 0; all other $\hat A_i^{\sigma}$ = 0 &
$\hat A_{\bbar}^{+}$, Im($X^{V*}_{LL}$) \\
\hline
\end{tabular}
\caption{NP parameter choices for each of the templates. 
\tm0, \tm1, \tm2, \tm3, \tm4 are the ones actually included in the fit. 
The template histograms have been generated with $10^6$ events each so that the 
statistical uncertainty originating from them is negligible and does not affect
the fit.}
\label{tab:templates}
\end{table}
\end{center}

\begin{center}
\begin{table}[ht]
\renewcommand{\arraystretch}{1.5}
\renewcommand{\tabcolsep}{0.2cm}
\begin{tabular}{|c|c|c|c|c|}
\hline 
Model & 
Parameter & 
Input Value &
Fit Result & 
$\chi^2$/d.o.f. \\
\hline \hline
\ex{C} & 
$\hat A_{c}^{+} + \hat A_{c}^{-} + \hat A_{\bar b}^{+} + \hat A_{\bar b}^{-} + \hat A_{b}^{+} + \hat A_{b}^{-}$ & 
64 &
65.7 $\pm$ 0.3 & 
1.2 \\
 &
$\hat A_{b}^{-} - \hat A_{b}^{+} - \hat A_{c}^{-} + \hat A_{c}^{+}$ &
32 &
29.5 $\pm$ 4.1 & \\
 &
Im($X^T_{LL}X^{S*}_{LL} + X^T_{RR}X^{S*}_{RR}$) &
4 &
3.1 $\pm$ 0.2 &
 \\
\hline  
\ex{D} & 
$\hat A_{c}^{+} + \hat A_{c}^{-} + \hat A_{\bar b}^{+} + \hat A_{\bar b}^{-} + \hat A_{b}^{+} + \hat A_{b}^{-}$ & 
77 &
77.6 $\pm$ 0.3 & 
1.3 \\
 &
$\hat A_{b}^{-} - \hat A_{b}^{+} - \hat A_{c}^{-} + \hat A_{c}^{+}$ &
67 &
62.8 $\pm$ 4.4 & \\
 &
Im($X^T_{LL}X^{S*}_{LL} + X^T_{RR}X^{S*}_{RR}$) &
3.5 &
2.8 $\pm$ 0.2 &
 \\
\hline  
\end{tabular}
\caption{Input values and fit results for the double differential
distribution in $\cos(\theta_5^*)$ and $\phi_1^{**}$.
The theoretical expression for the angular distribution is given in Eq.~(\ref{eq:TP1});
the actual fit is performed using templates, as described by Eq.~(\ref{eq:D}).}
\label{tab:fit}
\end{table}
\end{center}


\subsubsection{\texorpdfstring{$\langle \cos\theta_5^* \sin\phi_1^{**}\rangle$}{\textless ~cos(theta\_5)~sin(phi\_1)~ \textgreater}}
\label{sec:expecval_sim}

\begin{figure}[!htbp]
\includegraphics[scale=0.6]{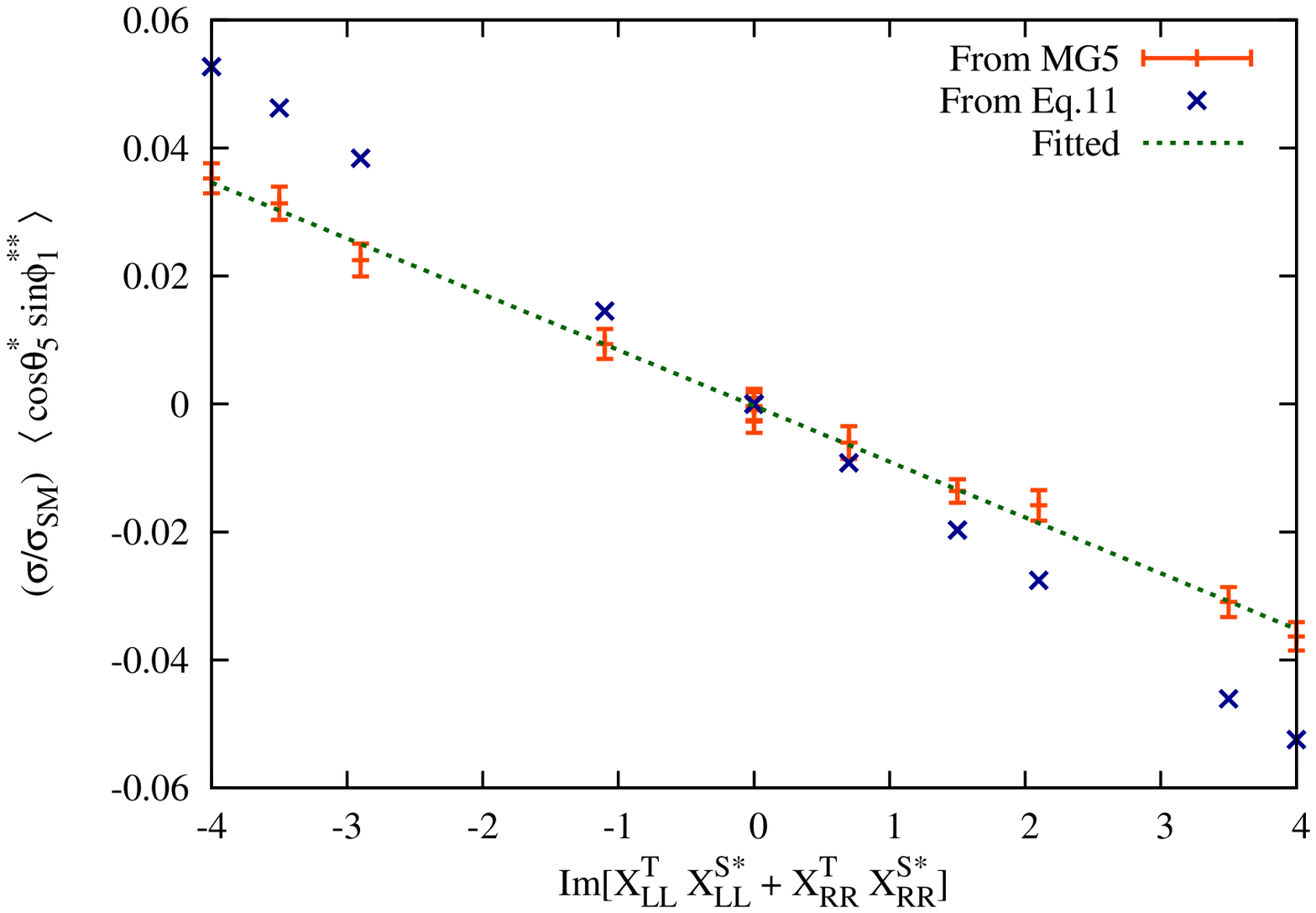}
\caption{Plot of 
$(\sigma / \sigma_{\rm SM})\,(\langle \cos\theta_5^* \sin\phi_1^{**}\rangle)$ 
as a function of Im[$X^T_{LL}X^{S*}_{LL}+X^T_{RR}X^{S*}_{RR}$] for various 
combinations of NP parameters.  As is evident from the plot, the simulated 
data has a linear dependence on Im[$X^T_{LL}X^{S*}_{LL}+X^T_{RR}X^{S*}_{RR}$], 
as in Eq.~(\ref{eq:expec_val}), even though Eq.~(\ref{eq:expec_val}) was 
derived under several simplifying assumptions.}
\label{fig:EVpoas}
\end{figure}

In Sec.~\ref{sec:expecval}, we saw that 
$\langle \cos\theta_5^* \sin\phi_1^{**}\rangle$
can be expressed as
\begin{eqnarray}
\langle \cos\theta_5^* \sin\phi_1^{**}\rangle 
\quad &&= \quad 
\frac{\sigma_{\mbox{\scriptsize SM}}}{\sigma} 
\;\; {\cal W} \;\; 
\mbox{Im}\!\left[X^T_{LL}X^{S*}_{LL}+X^T_{RR}X^{S*}_{RR}\right] \mspace{150mu}
\nonumber \\[2ex]
\text{where} \mspace{90mu} {\cal W} \quad &&= \quad 
\left(\frac{2\sqrt{2}G_F m_t^2 \; \langle \kappa(r) \rangle}
           {35 \left(1-\zeta_W^2\right)^2
\left(1+2\zeta_W^2\right)}\right) \,.
\nonumber
\end{eqnarray}
Using {\madg}, we generate several data-sets with different 
input values of Im[$X^T_{LL}X^{S*}_{LL}+X^T_{RR}X^{S*}_{RR}$].
We then calculate and plot 
$(\sigma / \sigma_{\rm SM})\,(\langle \cos\theta_5^* \sin\phi_1^{**}\rangle)$
for each data-set. These are shown as orange `$+$'s in Fig.~\ref{fig:EVpoas}. 
We also calculate and plot 
${\cal W}$\,Im[$X^T_{LL}X^{S*}_{LL}+X^T_{RR}X^{S*}_{RR}$]
using the input value of Im[$X^T_{LL}X^{S*}_{LL}+X^T_{RR}X^{S*}_{RR}$]
and the value of $\langle \kappa(r) \rangle$ obtained from the SM 
data-set\footnote{Since $\langle \kappa(r) \rangle$ depends only on the
$t \bar t$ production process, it is independent of NP.}.
These are the blue $\times$s in Fig.~\ref{fig:EVpoas}.
We see that, although the `$+$'s and `$\times$'s do not coincide,  
$(\sigma / \sigma_{\rm SM})\,(\langle \cos\theta_5^* \sin\phi_1^{**}\rangle)$
is, nonetheless, a linear function of 
Im[$X^T_{LL}X^{S*}_{LL}+X^T_{RR}X^{S*}_{RR}$] with zero intercept.

The linearity of the plot in Fig.~\ref{fig:EVpoas} has important ramifications. 
Firstly, we realize that a measurement of 
$\langle \cos\theta_5^* \sin\phi_1^{**}\rangle$ 
is, by itself, sufficient to indicate the 
presence of CP-violating new physics in $\tbbc$. 
Secondly, if such new physics does 
indeed exist in nature, then knowledge of the slope of the green dashed line 
in Fig.~\ref{fig:EVpoas}, along with
$\sigma_{\mbox{\scriptsize SM}}$, puts us in a position to directly 
extract the value of $\mbox{Im}\!\left[X^T_{LL}X^{S*}_{LL}+X^T_{RR}X^{S*}_{RR}\right]$
by simply measuring 
$\sigma$ and $\langle \cos\theta_5^* \sin\phi_1^{**}\rangle$.
Thirdly, while the PRA is sensitive to vectorial couplings 
(specifically Im($X^{V*}_{LL}$)), 
$\langle \cos\theta_5^* \sin\phi_1^{**}\rangle$ 
gives us a handle on scalar and tensorial NP couplings.


\subsubsection{\texorpdfstring{$\ACP$}{A\_TP}}
\label{sec:ACP_sim}

$\ACP$ is perhaps the simplest observable that can provide
an indication of the TP contributions due to NP.
Table~\ref{tab:ACP} shows the values of $\ACP$ obtained 
for the benchmark scenarios \ex{C} and \ex{D}, as well as the SM. 
It appears that $\ACP$ can prove to be an effective discriminator 
between SM and CP-violating NP. Of course, our analysis is a 
simple-minded one and the errors quoted are only statistical. 
Nevertheless, we feel that this is an observable worth experimental 
exploration, if only for its easy accessibility.  

\begin{center}
\begin{table}[ht]
\renewcommand{\arraystretch}{1.5}
\renewcommand{\tabcolsep}{0.2cm}
\begin{tabular}{|c|c|}
\hline 
Model & 
$\ACP$ \\
\hline \hline
\texttt{SM} & 
$0.004 \pm 0.004$ \\
\hline
\ex{C} & 
$-0.021 \pm 0.002$ \\
\hline
\ex{D} & 
$-0.015 \pm 0.002$ \\
\hline
\end{tabular}
\caption{Numerical results for the TP asymmetry defined in Eq.~(\ref{eq:ATP}) 
for the SM and two NP models.}
\label{tab:ACP}
\end{table}
\end{center}




\section{Feasibility}
\label{sec:feasibility}

The above analysis, and indeed those in Refs.~\cite{tNPLHC1,tNPLHC2},
is largely theoretical. On the whole, experimental considerations have
not been taken into account\footnote{There are two exceptions. We
  include a $b$-tagging efficiency in our estimate of the number of
  events produced after a certain number of years. And we include a
  kinematic cut to determine which of the two ${\bar b}$'s in the
  final state came from the $t$ and which came from the ${\bar t}$.}.
But this raises the question of feasibility: can the process
$\ggprocess$ even be seen\footnote{We remind the reader that although
  we refer to the process as arising from gluon fusion, our analysis
  also includes events coming from $q{\bar q}$ annihilation.}? While a
definitive answer cannot be given at this point, based on the
following discussion it appears that the chances are reasonably good
that the process can be observed \cite{JFA}.

As noted earlier, the LHC is essentially a top-quark factory. Thus,
even though $|V_{cb}|^2 = O(10^{-3})$, there should be many $\tbbc$
decays. The main difficulty is extracting the signal of this decay
from the very large background. To be specific, the signal of
$\ggprocess$ will involve three $b$ jets, one $c$ jet, one charged
lepton, and missing $E_T$. The dominant background is expected to be
$gg\to t\tbar\to\left(b\sbar c\right) \left(\bbar\ell\nubar\right)$,
which contains two $b$ jets, one $c$ jet, one light ($s$) jet, one
charged lepton, and missing $E_T$. The signal and background thus look
very similar -- the only difference is that one $b$ jet (signal) is
replaced by a light jet (background). Furthermore, the background is
roughly three orders of magnitude larger than the signal. Clearly the
analysis for extracting the signal won't be easy.

The key to differentiate the signal from background is to precisely
tag (i.e., identify) the $b$ jets and to distinguish them from
light-quark jets. This is done using properties such as the presence
of a secondary vertex inside the jet (with a high mass), and many
tracks with high impact parameters. $b$ tagging is discussed in a
recent note from the ATLAS Collaboration \cite{ATLASpub}. In Fig.~11
of this reference it is found that, for a $b$-tagging efficiency of
$\sim 65\%$, a rejection factor of $\sim 10^3$ can be obtained (these
numbers are relevant for Run 2 of the LHC). This leads to a
signal-to-background ratio approaching 1:1, as can be seen as follows.
Above we noted that, in searching for the $\tbbc$ signal, the dominant
background involves $t \to b\sbar c$. This means we expect roughly one
signal event for every $|V_{cs}|^2/|V_{cb}|^2 = 575$ \cite{pdg}
background events. Now, suppose there are 1000 signal events and hence
$\sim 575000$ background events.  The signal requires an additional
$b$ tag. A 65\% tagging efficiency leaves 650 signal events. On the
other hand, the rejection factor of $10^3$ leaves $\sim 575$
background events, for a signal-to-background ratio of about 1:1.  If
one can predict the background fairly precisely, then, just based on
this argument it should be possible to eventually observe a signal
over the background.

It is also likely to be necessary to tag the $c$ jet in order to
differentiate the signal from the large background. A charm tagger has
been developed by the ATLAS Collaboration \cite{charmtag}. For an
efficiency of 25\% in tagging $c$ jets, rejection factors of $\approx
100$ and $\approx 20$ are obtained for light and $b$ jets,
respectively.

Other important expected backgrounds are the associated production of
$t{\bar t}$ pairs with $b{\bar b}$ or $c{\bar c}$, producing four $b$
jets or two $b$ $+$ two $c$ jets. To deal with these, good $b$ and $c$
tagging will be necessary. These backgrounds could be reduced further
by searching for a peak in the mass of the $b{\bar b}c$ jets coming
from the top quark. This is non-trivial because it is necessary to
determine which of the three $b$ jets belongs to the other top quark
in the event, and so leads to a combinatorial background. Still, there
are tools to deal with this, such as reconstructing the whole event
with a kinematic fitter \cite{KLFitter}.

Admittedly, this is all speculative. A firm answer will only be
obtained when the experiment actually looks for $\ggprocess$. Its
observation will certainly require a good amount of data because the
signal efficiency will be reduced due to the requirement of three $b$
jets (tagging efficiency: $0.65^3 = 27$\%), as well as the hard cuts
necessary to see the signal above background. Still, given the number
of experimental handles (and the ingenuity of experimentalists), it
does appear that a measurement of $\ggprocess$ will be possible. Once
this is done, one can then apply the various proposed methods to
search for the presence of new physics.

\section{Conclusions}
\label{sec:conclusions}

This paper builds on the work done in
Refs.~\cite{kklrsw,tNPLHC1,tNPLHC2}, in which (NP)
contributions to $\tbbc$ were considered. 
Because this decay is
suppressed in the SM -- the amplitude is proportional to $V_{cb}$
($\sim$ 0.04) -- the NP effects could potentially be sizeable.
Reference~\cite{kklrsw} allows for all Lorentz structures, so that there
are ten possible dimension-6 NP operators that can contribute to
$\tbbc$. References~\cite{tNPLHC1,tNPLHC2} look primarily at CP-conserving NP
effects, and examine the prospects for their measurement at the LHC.
Here we perform a similar analysis, but for CP-violating effects.

At the LHC, single-top production is suppressed. $\tbbc$ must
therefore be studied within the context of $t \tbar$ pair
production. To be specific, we consider the semi-leptonic channel $gg
\to t (\to b \bbar c) \tbar (\to \bbar\ell\nubar)$. Here the
observation of a negatively-charged lepton indicates that it is the
$t$ that is undergoing the rare decay. 

We find that there are two types of CP-violating observables. The
first is the partial rate asymmetry (PRA), which compares the cross
section for $gg \to t (\to b \bbar c) \tbar (\to \bbar\ell\nubar)$ to
that for $gg \to \tbar (\to \bbar b \cbar) t (\to b\ellbar\nu)$. Now,
all CP-violating effects are due to the interference of two
amplitudes, and a nonzero PRA requires that these amplitudes have both
weak- and strong-phase differences. The NP strong phases are
negligible, but the SM $W$-mediated amplitude has a strong phase due
to the width of the $W$. Thus, the PRA arises from SM-NP interference,
which requires that the NP Lorentz structure be $(V-A)\times (V-A)$
(i.e., $\bbar\gamma_\mu P_L t \, \bar{c}\gamma^\mu P_L b$).  Despite
the suppression by $\Gamma_W/m_W$, a PRA whose magnitude is in excess
of 10\% is possible \cite{kklrsw}.

The second type of observable is a triple product (TP). A TP takes the
form $\vec v_1 \cdot (\vec v_2 \times \vec v_3)$ in the square of the
total amplitude of a decay process, where each $v_i$ is a spin or
momentum. The TP is odd under time reversal. Due to the presence of
strong phases, a truly CP-violating observable can be obtained only by
comparing the TP in the process with that in the CP-conjugate
process. In Ref.~\cite{kklrsw}, it was shown that, in the presence of
NP, one can generate a TP of the form
$\vec{s}_t\cdot\left(\vec{p}_{\overline{b}}\times \vec{p}_c\right)$ in
the decay $\tbbc$, where $\vec{s}_t$ is the spin of the top quark, and
$\vec{p}_i$ is the momentum of the particle $i$. However, this TP is
generated only through NP-NP interference, in which one of the NP
Lorentz structures is scalar ($S$), the other tensor ($T$).  And since
the NP strong phases are negligible, the TP is by itself a signal of
CP violation.  In the full process, $gg \to t (\to b \bbar c) \tbar
(\to \bbar\ell\nubar)$, one obtains information about $\vec{s}_t$ by
using the fact that, in $t \tbar$ production, the spins of the $t$ and
$\tbar$ are correlated \cite{spincorrexpt}.  Since $\vec{s}_\tbar$ is
related to the momenta of the decay products of the $\tbar$, the TP in
$\tbbc$ can be rewritten as a TP involving the final-state momenta of
the decay products of the $t$ and $\tbar$.
Furthermore, other TPs also appear, which involve initial state momenta.

The PRA and TP involve different combinations of NP parameters: the
PRA is due to SM-NP interference in which the NP is $(V-A)\times
(V-A)$, while the TP arises from the interference of $S$ and $T$ NP.
In order to see how well these observables can be used to detect the
presence of NP, and to measure the associated combinations of NP
parameters, we perform a Monte Carlo analysis using {\madg} along
with {\feynr}. This analysis follows that of Ref.~\cite{tNPLHC2}, and
includes (i) a method for distinguishing the $\bbar$'s coming from the
$t$ and $\tbar$ decays in order to identify the $\bbar$ in $\tbbc$,
(ii) the contribution to $t \tbar$ production from a $q {\bar q}$
initial state, and (iii) the PDFs for the initial-state partons. The
analysis is performed for an integrated luminosity corresponding to
$10^5$ SM events of the type $p p \to t \tbar \to (b \bbar c)
(\bbar\ell\nubar)$, which is expected to be achieved by the year 2030.

For the PRA, we find that, if it is large enough to be measured, there
is little difficulty in extracting the value of the $(V-A)\times
(V-A)$ NP parameter. The PRA is therefore an excellent observable for
measuring one type of CP-violating NP in $\tbbc$.

For the TP, the analysis is more complicated. We find three
observables that can be used to probe the TP. One involves both
CP-conserving and CP-violating NP, the other two are purely
CP-violating. In all three cases, it is straightforward to obtain
statistically-significant evidence that CP-violating NP is present.
We examine two methods for extracting the value of the combination of 
NP parameters responsible for the TP. The first involves a weighted fitting 
of histograms (described in Sec.~\ref{sec:angdist_sim}), and does not lead to a 
very accurate extraction of the desired parameter. This is related to the 
fact that the CP-violating parameter in the TP is due to a particular 
combination of the operators introduced in the Lagrangian. Each of these 
operators also leads to CP-conserving contributions. Subtracting out the 
CP-conserving part from the histograms, while retaining the CP-violating part, 
should ideally include a careful consideration of the correlations between 
these two contributions. However, these have been ignored in our relatively 
simple-minded analysis.

Interestingly, these kinds of complications can be simply avoided by 
adopting a graphical method (discussed in Sec.~\ref{sec:expecval_sim}), which 
fares much better in the task of extracting the relevant CP-violating 
combination of NP parameters.


\bigskip
\noindent
{\bf Acknowledgments:} The authors wish to thank the \textsc{MadGraph}
and {\feynr} Teams for extensive discussions, S.~Judge for
collaboration at an early stage of this work, K.~Constantine for
helpful discussions and R.~Godbole for useful comments. We are
grateful to J.-F. Arguin for his detailed explanation about how to
measure $\ggprocess$, as well as its feasibility.
This work was financially supported by NSERC of Canada (BB, DL, PS)
and DST, India (PS).
This work has been partially supported by CONICET (AS).  
The work of KK was supported by the U.S.\ National Science Foundation 
under Grant PHY--1215785. KK also acknowledges sabbatical support from 
Taylor University.
PS would like to thank IRC, University of Delhi and 
RECAPP, Harish-Chandra Research Institute for hospitality 
and computational facilities during different stages of this work.

\newpage


\begin{center}
\section*{APPENDIX}       
\end{center}

\begin{appendices}
\renewcommand\thesubsection{\arabic{subsection}}
\renewcommand{\theequation}{\Alph{section}.\arabic{equation}}

\setcounter{equation}{0}
\section{\texorpdfstring{\boldmath Cross section for $\ggprocess$}{Cross section for gluon fusion}}
\label{sec:app1}

\subsection{Differential cross-section}
\label{sec:csec_diff}

\noindent
We have
\begin{eqnarray}
  d\sigma\left(\ggprocess\right) =
    \left({\cal B}_{\mbox{\scriptsize non-TP}}
       + {\cal B}_{\mbox{\scriptsize TP}}\right)d\lambda ~,
\label{eq:dsigma2}
\end{eqnarray}
where
\begin{eqnarray}
\label{eq:Bnon-TP2}
{\cal B}_{\mbox{\scriptsize non-TP}} & = &
\sum_{i,\sigma} A_i^\sigma A_\ell
    \Big\{
    -\frac{p_i\!\cdot\! p_t \,p_\ell\!\cdot\! p_{\overline{t}}}{m_t^2}
    \left[f(r,z)\!+\!\xi^\sigma
\left(r^4\left(z^4-2\right)\!+\!1\right)\right] 
       \!-\! \xi^\sigma p_i\!\cdot\! p_\ell \, g(r,z) \nonumber\\
    && ~~~~~~~~~~~~~~-\frac{\left(r^2-1\right)
         \left[r^2\left(z^2-2\right)+1\right]
         \xi^\sigma}{2 m_t^2}
     \left(p_i\!\cdot\!Q \,Q\!\cdot\!p_\ell
         + p_i\!\cdot\!P_t \,P_t\!\cdot\!p_\ell\right)\nonumber\\
    && ~~~~~~~~~~~~~~-
    \frac{r^2(r^2-1)(z^2-1)\xi^\sigma}{2 m_t^2}\big[
      p_i\!\cdot\! P_g \left(
       P_g\!\cdot\! p_\ell - Q\!\cdot\! p_\ell\,rz
      \right) \nonumber\\
     && ~~~~~~~~~~~~~~~~~~~~~~~~~~~~~~~~~~~~~~~~~~+
      p_i\!\cdot\! Q \left(
       P_g\!\cdot\! p_\ell\,rz - Q\!\cdot\! p_\ell
      \right)
      \big]
    \Big\} ~, \\
{\cal B}_{\mbox{\scriptsize TP}} & = &
    16 A_\ell
\,\mbox{Im}\left(X^T_{LL}X^{S*}_{LL}+X^T_{RR}X^{S*}_{RR}\right)
    \Big\{ 
      -g(r,z)\epsilon\left(p_b, p_{\overline{b}},p_c,p_\ell\right)
      \nonumber\\
    && - \frac{\left(r^2-1\right)p_\ell\!\cdot\!p_{\overline{t}}}
      {m_t^2}\left[r^2\left(z^2-2\right)+1\right]
        \epsilon\left(p_b, p_{\overline{b}},p_c,Q\right)\nonumber\\
    && -\frac{r^2(r^2-1)(z^2-1)}{2 m_t^2}\big[
      \left(
       P_g\!\cdot\! p_\ell - Q\!\cdot\! p_\ell\,rz
      \right)\epsilon\left(p_b, p_{\overline{b}},p_c,P_g\right)
\nonumber\\
    && ~~~~~~~~~~~~~~~~~~~~~~~~+
      \left(
       P_g\!\cdot\! p_\ell\,rz - Q\!\cdot\! p_\ell
      \right)\epsilon\left(p_b, p_{\overline{b}},p_c,Q\right)
      \big]   
    \Big\} ~,
    \label{eq:BTP2}
\end{eqnarray}
and
\begin{eqnarray}
  d\lambda & = &
    \frac{\alpha_S^2 \,G_F^4 V_{tb}^4V_{cb}^2 \left(1-r^2\right) r }
      {4 \left(4\pi\right)^{10} \Gamma_t^2 \,m_t^2}
    \left(1-\frac{M_2^2}{m_t^2}\right)
    \left(1-\frac{M_5^2}{m_t^2}\right)
    \frac{(9r^2z^2+7)}{(r^2z^2-1)^2} \nonumber\\
    && \times dM_2^2 \,dM_5^2\,
d\Omega_1^{**}\,d\Omega_2^{*}\,d\Omega_4^{**}\,d\Omega_5^{*}\,d\Omega_t
~.
\label{eq:dlambda}
\end{eqnarray}
In the above, the $p_i$ are the momenta of the final-state quarks
coming from the top decay (i.e., $b$, $\bbar$ and $c$).  Also, $\sigma = \pm$,
$\xi^\pm = \pm 1$, and
$\epsilon(p_1, p_2, p_3, p_4) \equiv \epsilon^{\alpha\beta\gamma\delta}p_{1\alpha}p_{2\beta}p_{3\gamma}p_{4\delta}$,
with $\epsilon^{0123} = +1$.  Furthermore,
\begin{eqnarray}
  & P_t \equiv p_t-p_{\tbar} ~,~~
  Q \equiv q_1+q_2 = p_t+p_{\tbar} ~,~~
  P_g \equiv q_1-q_2 ~, & \nn\\
  & r \equiv \sqrt{1 - 4 m_t^2/Q^2} ~,~~  
  z \equiv - P_t \cdot P_g/(r Q^2) ~, &
\label{eq:Pt}
\end{eqnarray}
where $p_t$ and $p_{\tbar}$ are the $t$ and $\tbar$ momenta, and $q_1$
and $q_2$ are the momenta of the initial gluons, and 
\begin{eqnarray}
  f(r,z) & = & z^4r^4+2r^2z^2\left(1-r^2\right)+2r^4-2r^2-1 ~,
  \label{eq:f} \\
  g(r,z) & = & r^4\left(z^4 -2z^2+2\right)-2r^2+1 ~.
  \label{eq:g}
\end{eqnarray}
$A_{\bbar}^+$ is defined as
\begin{eqnarray}
  A_{\bbar}^+ = \left(p_t-p_{\bbar}\right)^2
  \Big[m_W^4\left|G_T\right|^2
  +4 m_W^2\mbox{Re}\left(G_T X^{V*}_{LL}\right)
+\hat{A}_{\bbar}^+\Big] ~,
\label{eq:Abbar}
\end{eqnarray}
where $G_T \equiv G_T(q^2) = (q^2 - M_W^2 + i \Gamma_W M_W)^{-1}$ 
and $q^2 = 2 \,p_{\bbar}\cdot p_c$. The remaining $A_i^\sigma$ are
defined as
\begin{eqnarray}
  A_i^\sigma = \left(p_t-p_{i}\right)^2 \hat{A}_{i}^\sigma ~,
  ~~~~~~~~~~~~~~~~~~~~~~~
  \mbox{(all $i,\sigma$, except $i=\bbar$, $\sigma = +$).}
\end{eqnarray}
In the above,
\begin{eqnarray}
  \hat{A}_{\bbar}^+ \;\; &=& \;\; 4 \left|X^{V}_{LL}\right|^2
  -8 \,\mbox{Re}\left(X^T_{LL}X^{S*}_{LL}\right)+32 \left|X^T_{LL}\right|^2 ~,
  \mspace{100mu} \nn\\
  \hat{A}_{\bbar}^- \;\; &=& \;\;
    4\left|X^{V}_{RR}\right|^2
    -8 \,\mbox{Re}\left(X^T_{RR}X^{S*}_{RR}\right)+32 \left|X^T_{RR}\right|^2 ~, \nn\\
  \hat{A}_{b}^+ \;\; &=& \;\; 
    \left|X^{S}_{LL}\right|^2+\left|X^{S}_{LR}\right|^2
    -16\left|X^{T}_{LL}\right|^2 ~, \nn\\
  \hat{A}_{b}^- \;\; &=& \;\; 
    \left|X^{S}_{RR}\right|^2+\left|X^{S}_{RL}\right|^2
    -16\left|X^{T}_{RR}\right|^2 ~,\nn\\
  \hat{A}_{c}^+ \;\; &=& \;\; 
    4\left|X^{V}_{LR}\right|^2
    +8 \,\mbox{Re}\left(X^T_{LL}X^{S*}_{LL}\right)+32\left|X^T_{LL}\right|^2 ~, \nn\\
  \hat{A}_{c}^- \;\; &=& \;\; 
    4\left|X^{V}_{RL}\right|^2
    +8 \,\mbox{Re}\left(X^T_{RR}X^{S*}_{RR}\right)+32\left|X^T_{RR}\right|^2 ~.
\label{eq:Ahatdefs}
\end{eqnarray}


\subsection{Integrated cross-section}
\label{sec:csec_intg}

\noindent
The tree-level SM cross section for $\ggprocess$ is
\begin{eqnarray}
\sigma_{\mbox{\scriptsize SM}}
\quad &&\equiv \quad
\sigma\left(\ggprocess\right)\big|_{\mbox{\scriptsize SM}} \nonumber
\\[2ex]
\quad &&= \quad
\sigma \!\left(gg\to t\tbar\right) 
\mbox{BR}\!\left.\left(t\to b\bbar c \right)\right|_{\mbox{\scriptsize
SM}}
\mbox{BR}\!\left(\tbar\to \bbar\ell\nubar\right),
\label{eq:sigmaSM}
\end{eqnarray}
where BR$\left.\left(t\to b\bbar c\right)
\right|_{\mbox{\scriptsize SM}}\! = \!V_{tb}^2V_{cb}^2/3$,
BR$\left(\tbar\to \bbar\ell\nubar\right) \! =\!  V_{tb}^2/9$ and
\begin{eqnarray}
\sigma \left(gg\to t\tbar\right) 
     \!& = & \!\frac{\pi \alpha_S^2 (1-r^2)}{192 m_t^2}
     \left[r(31 r^2-59) + 2 (r^4-18 r^2+33)\tanh^{-1}(r)\right]. \mspace{50mu}
\label{eq:sigggttbar}
\end{eqnarray}

\noindent
After the inclusion of new physics,
\begin{eqnarray}
\sigma_{\mbox{\scriptsize SM+NP}}
\quad &&\equiv \quad
  \sigma\left(\ggprocess\right) \big|_{\mbox{\scriptsize SM+NP}} \nn 
\\[2ex]
  &&= \quad \sigma_{\mbox{\scriptsize SM}}
    \Bigg\{
    1+ \frac{4\Gamma_W}{m_W}\mbox{Im}\left(X^{V*}_{LL}\right)
  + \frac{3 G_F m_t^2}
     {4\sqrt{2}\pi^2 \left(1-\zeta_W^2\right)^2
      \left(1+2\zeta_W^2\right)}
    \sum_{i,\sigma} \hat{A}_i^\sigma\Bigg\}. \mspace{70mu} 
\label{eq:sigggttbar_int}
\end{eqnarray}



\setcounter{equation}{0}
\section{\texorpdfstring{\boldmath Cross section for $\qqbarprocess$}{Cross section for quark-antiquark annihilation}}
\label{sec:app2}

\subsection{Differential cross-section}
\label{sec:csec_diff_qq}

\noindent
The expression for the differential cross section for 
$\qqbarprocess$ 
can be determined using the approach
described in the appendix of Ref.~\cite{tNPLHC1} (see also Ref.~\cite{Kawasaki}).  For
the $q\overline{q}$ case we find
\begin{eqnarray}
  d\sigma\left(\qqbarprocess\right) =
    \left({\cal B}_{\mbox{\scriptsize non-TP}}^{q\overline{q}}
       + {\cal B}_{\mbox{\scriptsize TP}}^{q\overline{q}}\right)d\lambda_{q\overline{q}} ~,
\label{eq:dsigma2_qq}
\end{eqnarray}
where
\begin{eqnarray}
\label{eq:Bnon-TP2_qq}
{\cal B}_{\mbox{\scriptsize non-TP}}^{q\overline{q}} & = &
\sum_{i,\sigma} A_i^\sigma A_\ell
    \Bigg\{
    \frac{p_i\!\cdot\! p_t \,p_\ell\!\cdot\! p_{\overline{t}}}{m_t^2}
    \left[
      2+r^2(z^2-1) +r^2(1+z^2)\xi^\sigma
        \right] \nonumber\\[5pt]
      && ~~~~~~~~~~~~~~
       - p_i\!\cdot\! p_\ell \, r^2(1-z^2)\xi^\sigma  \nonumber\\[10pt]
    && ~~~~~~~~~~~~~~-\frac{
         \left(1-r^2\right)
         \xi^\sigma}{2 m_t^2}
     \big[p_i\!\cdot\!Q \,Q\!\cdot\!p_\ell
     + p_i\!\cdot\!P_t \,P_t\!\cdot\!p_\ell
      \nonumber\\
    && ~~~~~~~~~~~~~~~~~~~~~~~~~~~~~~~~+
      p_i\!\cdot\! P_g \left(
       P_g\!\cdot\! p_\ell - Q\!\cdot\! p_\ell\,rz
      \right) \nonumber\\
     && ~~~~~~~~~~~~~~~~~~~~~~~~~~~~~~~~+
      p_i\!\cdot\! Q \left(
       P_g\!\cdot\! p_\ell\,rz - Q\!\cdot\! p_\ell
      \right)
      \big]
    \Bigg\} ~, \\ 
{\cal B}_{\mbox{\scriptsize TP}}^{q\overline{q}} & = &
    16 A_\ell
\,\mbox{Im}\left(X^T_{LL}X^{S*}_{LL}+X^T_{RR}X^{S*}_{RR}\right)
    \Big\{ 
      -r^2(1-z^2)\epsilon\left(p_b, p_{\overline{b}},p_c,p_\ell\right)
      \nonumber\\
    && ~~~~~~~~~~- \frac{p_\ell\!\cdot\!p_{\overline{t}}}
      {m_t^2}\left(1-r^2\right)
        \epsilon\left(p_b, p_{\overline{b}},p_c,Q\right)\nonumber\\
    && ~~~~~~~~~~-\frac{(1-r^2)}{2 m_t^2}\big[
      \left(
       P_g\!\cdot\! p_\ell - Q\!\cdot\! p_\ell\,rz
      \right)\epsilon\left(p_b, p_{\overline{b}},p_c,P_g\right)
\nonumber\\
    && ~~~~~~~~~~~~~~~~~~~~~~~~+
      \left(
       P_g\!\cdot\! p_\ell\,rz - Q\!\cdot\! p_\ell
      \right)\epsilon\left(p_b, p_{\overline{b}},p_c,Q\right)
      \big]   
    \Big\} ~,
    \label{eq:BTP2_qq}
\end{eqnarray}
and
\begin{eqnarray}
  d\lambda_{q\overline{q}} & = &
    \frac{8\,\alpha_S^2 \,G_F^4 V_{tb}^4V_{cb}^2 \left(1-r^2\right) r }
         {3 \left(4\pi\right)^{10} \Gamma_t^2 \,m_t^2}
    \left(1-\frac{M_2^2}{m_t^2}\right)
    \left(1-\frac{M_5^2}{m_t^2}\right)
    \nonumber\\
    && \times dM_2^2 \,dM_5^2\,
d\Omega_1^{**}\,d\Omega_2^{*}\,d\Omega_4^{**}\,d\Omega_5^{*}\,d\Omega_t
~.
\label{eq:dlambda_qq}
\end{eqnarray}
In the above expressions, $P_g$, $Q$, $P_t$, $r$ and $z$ are defined as in
Eq.~(\ref{eq:Pt}), except that $q_1$ and $q_2$ are now
the momenta of the $q$ and $\overline{q}$, respectively.
The above expressions can be integrated to
determine any differential cross sections that are of interest.
Comparison with the analogous expressions that we had derived for the
gluon fusion case shows that the overall structure of the two sets of
expressions is very similar.  The main differences are in the
functions of $r$ and $z$ that multiply the various terms.  


\subsection{Integrated cross-section}
\label{sec:csec_intg_qq}

\noindent
The tree-level SM cross section for $\qqbarprocess$ is
\begin{eqnarray}
\sigma_{\mbox{\scriptsize SM}}^{q\overline{q}}
\quad &&\equiv \quad
\sigma\left(\qqbarprocess\right)\big|_{\mbox{\scriptsize SM}} \nonumber
\\[2ex]
\quad &&= \quad
\sigma \!\left(q\overline{q} \to t\tbar\right) 
\mbox{BR}\!\left.\left(t\to b\bbar c \right)\right|_{\mbox{\scriptsize
SM}}
\mbox{BR}\!\left(\tbar\to \bbar\ell\nubar\right),
\label{eq:sigmaSM_qq}
\end{eqnarray}
where BR$\left.\left(t\to b\bbar c\right)
\right|_{\mbox{\scriptsize SM}}\! = \!V_{tb}^2V_{cb}^2/3$,
BR$\left(\tbar\to \bbar\ell\nubar\right) \! =\!  V_{tb}^2/9$ and
\begin{eqnarray}
\sigma \left(q\overline{q}\to t\tbar\right) 
     \!& = & \!\frac{\pi \alpha_S^2 (1-r^2)(3-r^2)r}{27 m_t^2}.
\label{eq:sigqqbarttbar_qq}
\end{eqnarray}

\noindent
After the inclusion of new physics, 
$\sigma_{\mbox{\scriptsize SM+NP}}^{q\overline{q}}$ assumes the same form as
Eq.~(\ref{eq:sigggttbar_int}) with 
$\sigma_{\mbox{\scriptsize SM}}$ replaced by 
$\sigma_{\mbox{\scriptsize SM}}^{q\overline{q}}$.


\subsection{Angular distribution}
\label{sec:angdist_qq}

\noindent
Integrating the differential cross section over all phase space variables
except for the angles $\theta_5^*$ and $\phi_1^{**}$ yields
\begin{eqnarray}
  \frac{d\sigma_{q \bar q}}{d\!\cos\theta_5^* \,d\phi_1^{**}}
  &=& \frac{\sigma_{\mbox{\scriptsize SM}}^{q \bar q}}{4\pi} \Bigg\{ 1 +
    \frac{4\Gamma_W}{m_W}\mbox{Im}\left(X^{V*}_{LL}\right) 
    \; + \; \frac{3 G_F m_t^2}{4\sqrt{2}\pi^2 \left(1-\zeta_W^2\right)^2
      \left(1+2\zeta_W^2\right)} \Bigg(\sum_{i,\sigma}\hat{A}{_i^\sigma} \nonumber \\
    && \mspace{60mu} +\; \frac{2\pi^2\eta(r)}{35}\Big[\cos\theta_5^*\cos\phi_1^{**}
         \left(\hat{A}_b^--\hat{A}_b^+-\hat{A}_c^-+\hat{A}_c^+\right)\nonumber\\
 && \mspace{60mu} + \; 
 16 \cos\theta_5^*\sin\phi_1^{**} \mbox{Im}\!\left[X^T_{LL}X^{S*}_{LL}+X^T_{RR}X^{S*}_{RR}\right]
         \Big]\Bigg)\Bigg\},
\label{eq:TP1_qq}
\end{eqnarray}
where 
\begin{eqnarray}
  \eta(r) = \frac{(1+r^2)}{(3-r^2)}~,
\end{eqnarray}
with $r=\sqrt{1-4m_t^2/Q^2}$ and $Q\equiv q_1+q_2 = p_t+p_{\overline{t}}$.

\noindent
Comparing Eq.~(\ref{eq:TP1_qq}) with the analogous expression from the
gluon fusion case, we see that the current expression can be obtained
from the former one by the replacements $\kappa(r)\to \eta(r)$ and
$\sigma_{\mbox{\scriptsize SM}}\to \sigma_{\mbox{\scriptsize
    SM}}^{q\overline{q}}$.  What this means is that the angular
dependence, including its dependence on the NP parameters, is
practically identical in the $gg$ and $q\overline{q}$ cases.  What is
different in the two cases is the relative size and possibly the
sign of the angular terms compared to the constant terms; these are 
determined by $\kappa(r)$ in the $gg$ case and 
$\eta(r)$ in the $q\overline{q}$ case.



\newpage

\setcounter{equation}{0}
\section{\texorpdfstring{\boldmath Choice of $X^I_{AB}$ for Benchmark Scenarios}{Choice of Benchmark Scenarios}}
\label{sec:app3}

\begin{center}
\begin{table}[ht]
\renewcommand{\arraystretch}{1.5} \renewcommand{\tabcolsep}{0.2cm}
\begin{tabular}{|c|l|l|c|}
\hline 
Test Case & 
\multicolumn{1}{c|}{$X^I_{AB}$} & 
\multicolumn{1}{c|}{$\hat A_i^{\sigma}$} &
$\sigma / \sigma_{SM}$ \\
\hline \hline
\ex{A} & 
$X^V_{LL}$ = 3$i$ ; 
$X^V_{RR}$ = 1 ; &
$\hat A_{\bbar}^{+}$ = 36 ; 
$\hat A_{\bbar}^{-}$ = 4 ; &
2.3 \\
 &
$X^V_{LR}$ = 2$i$ ; 
$X^V_{RL}$ = 0 ; &
$\hat A_{b}^{+}$ = 17 ; 
$\hat A_{b}^{-}$ = 0 ; &
\\
 &
$X^S_{LL}$ = 4$i$ ; 
$X^S_{RR}$ = 0 ; &
$\hat A_{c}^{+}$ = 16 ; 
$\hat A_{c}^{-}$ = 0 ;  &
\\
 &
 $X^S_{LR}$ = 1 ; 
 $X^S_{RL}$ = 0 ; &
 Im($X^T_{LL}X^{S*}_{LL}+X^T_{RR}X^{S*}_{RR}$) = 0 & 
\\
 &
 $X^T_{LL}$ = 0 ; 
 $X^T_{RR}$ = 0 &
 & 
\\
\hline
\ex{B} & 
$X^V_{LL}$ = 2$i$ ; 
$X^V_{RR}$ = 2$i$ ; & 
$\hat A_{\bbar}^{+}$ = 26.08 ; 
$\hat A_{\bbar}^{-}$ = 24 ; &
3.2 \\
 &
$X^V_{LR}$ = 2$i$ ; 
$X^V_{RL}$ = 2 ; &
$\hat A_{b}^{+}$ = 16.56 ; 
$\hat A_{b}^{-}$ = 12 ; &
\\
 &
$X^S_{LL}$ = 3 + 3$i$ ; 
$X^S_{RR}$ = 4$i$ ; &
$\hat A_{c}^{+}$ = 11.68 ; 
$\hat A_{c}^{-}$ = 24 ;  &
\\
 &
$X^S_{LR}$ = 0 ; 
$X^S_{RL}$ = 0 ; &
Im($X^T_{LL}X^{S*}_{LL}+X^T_{RR}X^{S*}_{RR}$) = -2.9 & \\
 &
$X^T_{LL}$ = -0.3$i$ ; 
$X^T_{RR}$ = 0.5 &
 &
\\ 
\hline
\ex{C} & 
$X^V_{LL}$ = 0 ; 
$X^V_{RR}$ = 0 ; &
$\hat A_{\bbar}^{+}$ = 32 ; 
$\hat A_{\bbar}^{-}$ = 0 ; &
2.4 \\
  &
$X^V_{LR}$ = 0 ; 
$X^V_{RL}$ = 0 ;&
$\hat A_{b}^{+}$ = 0 ; 
$\hat A_{b}^{-}$ = 0 ; &
\\
 &
$X^S_{LL}$ = 4 ; 
$X^S_{RR}$ = 0 ; &
$\hat A_{c}^{+}$ = 32 ; 
$\hat A_{c}^{-}$ = 0 ;  &
 \\
 &
$X^S_{LR}$ = 0 ; 
$X^S_{RL}$ = 0 ; &
Im($X^T_{LL}X^{S*}_{LL}+X^T_{RR}X^{S*}_{RR}$) = 4 & \\
 &
$X^T_{LL}$ = $i$ ; 
$X^T_{RR}$ = 0 &
 &
\\ 
\hline
\ex{D} & 
$X^V_{LL}$ = 0 ; 
$X^V_{RR}$ = 0 ; &
$\hat A_{\bbar}^{+}$ = 16 ; 
$\hat A_{\bbar}^{-}$ = 0 ; &
2.7 \\
  &
$X^V_{LR}$ = 0 ; 
$X^V_{RL}$ = 0 ;&
$\hat A_{b}^{+}$ = -3 ; 
$\hat A_{b}^{-}$ = 0 ; &
\\
 &
$X^S_{LL}$ = 4 + $i$ ; 
$X^S_{RR}$ = 0 ; &
$\hat A_{c}^{+}$ = 64 ; 
$\hat A_{c}^{-}$ = 0 ;  &
 \\
 &
$X^S_{LR}$ = 0 ; 
$X^S_{RL}$ = 0 ; &
Im($X^T_{LL}X^{S*}_{LL}+X^T_{RR}X^{S*}_{RR}$) = 3.5 & \\
 &
$X^T_{LL}$ = 0.5 + $i$ ; 
$X^T_{RR}$ = 0 &
 &
\\ 
\hline
\end{tabular}
\caption{Input values of the NP parameters for the four test cases.
The last column shows how the total cross section $\sigma$ is affected 
in each of the test cases. The values quoted are for 
$pp \to t\tbar\to\left(b\bbar c\right) \left(\bbar\ell\nubar\right)$.}
\label{tab:NPcases}
\end{table}
\end{center}


\end{appendices}


\newpage





\end{document}